\documentclass[tighten, twocolumn, trackchanges, 12pt]{aastex62}
\def\apjl{{ApJ}}                
\def\apjs{{ApJS}}               
\def\ao{{Appl.~Opt.}}           
\def\aap{{A\&A}}                

\usepackage{listings}
\usepackage{color}

\definecolor{dkgreen}{rgb}{0,0.6,0}
\definecolor{gray}{rgb}{0.5,0.5,0.5}
\definecolor{mauve}{rgb}{0.58,0,0.82}

\usepackage{graphicx}
\usepackage{hyperref}  
\begin{document}


\title[A Gaia DR2 Mock Stellar Catalog]{A Gaia DR2 Mock Stellar Catalog}

\correspondingauthor{Jan Rybizki}
\email{rybizki@mpia.de}

\author[0000-0002-0993-6089]{Jan Rybizki}
\affil{Max Planck Institute for Astronomy,
	K\"onigstuhl 17, D-69117 Heidelberg, Germany}

\author{Markus Demleitner}
\affil{Astronomisches Rechen-Institut, Zentrum f{\"u}r Astronomie der Universit{\"a}t Heidelberg, M{\"o}nchhofstrasse 12-14, D-69120 Heidelberg, Germany}

\author[0000-0001-9256-5516]{Morgan Fouesneau}
\affil{Max Planck Institute for Astronomy,
	K\"onigstuhl 17, D-69117 Heidelberg, Germany}

\author{Coryn Bailer-Jones}
\affil{Max Planck Institute for Astronomy,
	K\"onigstuhl 17, D-69117 Heidelberg, Germany}
\author[0000-0003-4996-9069]{Hans-Walter Rix}
\affil{Max Planck Institute for Astronomy,
	K\"onigstuhl 17, D-69117 Heidelberg, Germany}
\author[0000-0002-6274-6612]{Ren\'{e} Andrae}
\affil{Max Planck Institute for Astronomy,
	K\"onigstuhl 17, D-69117 Heidelberg, Germany}

\vspace{10pt}

\begin{abstract}
We present a mock catalog of Milky Way stars, matching in volume and depth the content of the Gaia data release 2 (GDR2).
We generated our catalog using \texttt{Galaxia}, a tool to sample stars from a Besan\c{c}on Galactic model, together with a realistic 3D dust extinction map. 
The catalog mimicks the complete GDR2 data model and contains most of the entries in the Gaia source catalog: 5-parameter astrometry, 3-band photometry, radial velocities, stellar parameters, and associated scaled nominal uncertainty estimates.  In addition, we supplemented the catalog with extinctions and photometry for non-Gaia bands.
This catalog can be used to prepare GDR2 queries in a realistic runtime environment, and it can serve as a Galactic model against which to compare the actual GDR2 data in the space of observables.
The catalog is hosted through the virtual observatory GAVO's Heidelberg data center\footnote{\url{http://dc.g-vo.org/tableinfo/gdr2mock.main}} service and thus can be queried using \texttt{ADQL} as for GDR2 data.
\end{abstract}
\keywords{Galaxy: stellar content; Astrometry; Catalogs}
%
%
%
%
%

\section{Introduction}
Gaia \citep{2016A&A...595A...1G} is an ongoing ESA astrometric space mission about to deliver positions, parallaxes, proper motions, and three photometric bands for a set of $\sim$1.4\,billion sources across the whole sky with its second data release (GDR2; \citealt{DR2-DPACP-51}). This dataset will also provide effective temperatures, luminosities, extinction estimates and radial velocity measurements for a substantial subset of those plus some other data products. This vast amount of data will be a practical challenge to explore and should usher the community into a new regime in Galactic stellar astronomy, where well-designed \texttt{ADQL}\footnote{ADQL = astronomical data query language} queries become a common tool to obtain manageable data sets from hosting services like the Virtual Observatory \citep[VO;][]{2014ASPC..485..309D}.

To help prepare the scientific community for this phase change, we present in this paper a mock catalog that contains the prospective GDR2 stellar content. A first mock dataset of Gaia data has long been available, the so-called Gaia Universe Model \citep[GUMS;][]{2012A&A...543A.100R}. However, the primary goal of that catalog was to provide simulations to the data processing consortium (DPAC). Hence, its design does not offer the same capabilities as our GDR2 mock catalog. In addition to an improved 3D extinction map which results in a slightly larger starcount (i.e. $\sim$1.1\,billion stars compared to $\sim$1.0\,bn in GUMS for stars brighter than $G=20$ and a total starcount of $\sim$1.6\,bn for our complete catalog down to $G=20.7$), the main difference by construction is that this catalog fully mimicks the GDR2 format. This enables GDR2 users to test their \texttt{ADQL} queries and helps with their science analysis (e.g. selection function).

\medskip
Our catalog is accessible online, most easily via \texttt{topcat} exploiting the VO table access protocol (TAP) service from GAVO\footnote{\url{http://dc.g-vo.org/tap}} where the catalog is referenced under \texttt{gdr2mock.main}

\section{catalog generation}

Our catalog is based on a chemo-dynamical model Milky Way, \texttt{Galaxia} \citep{2011ApJ...730....3S}, 
which we associated with a 3D dust extinction model before generating photometric observables. 
The following subsections outline the steps in this mock data set generation.\footnote{Part of the routines we used can be retrieved from \url{https://github.com/jan-rybizki/Galaxia_wrap}.}

\subsection{The Galaxia Model}\label{sec:galaxia}

\texttt{Galaxia} is a tool that allows one to sample stars from the Besan\c{c}on Galactic model \citep{2003A&A...409..523R}, using a specific set of stellar isochrones to obtain their astrophysical parameters. The Galactic warp was switched on during the simulations and the solar zero-point was set to $(X,Y,Z) = (-8.0,0.0,0.015)\,\mathrm{kpc}$ and the velocities to $(U,V,W) = (11.1,239.08,7.25)\,\frac{\mathrm{km}}{\mathrm{s}}$.  
Transformations from phase-space to observable coordinates on the sky (\texttt{ra}, \texttt{dec}, \texttt{pm\_ra\_cosdec}, \texttt{pm\_dec} and \texttt{radial\_velocity}) were done using astropy\footnote{\url{http://www.astropy.org}} \citep{2018arXiv180102634T}.
And we used the latest {PARSEC} isochrones\footnote{PARSEC = Padova Trieste evolution code (including the pre-main sequence phase); \url{http://stev.oapd.inaf.it/cgi-bin/cmd}} -- \texttt{PARSEC v1.2S+ COLIBRI PR16} \citep{2012MNRAS.427..127B,2017ApJ...835...77M,2016ApJ...822...73R,2013MNRAS.434..488M} -- which also provide photometric values for each star using the nominal Gaia DR1 photometric bands G, BP, and RP \citep{2010A&A...523A..48J}. GDR2 passbands where not available during the construction of this catalog.

At this stage, we were already able to account for the magnitude limit of Gaia and only selected stars with apparent magnitude brighter than $G=20.7$\,mag, which preliminarily resulted in over six billion sources.

\begin{figure}
	\includegraphics[width=\linewidth]{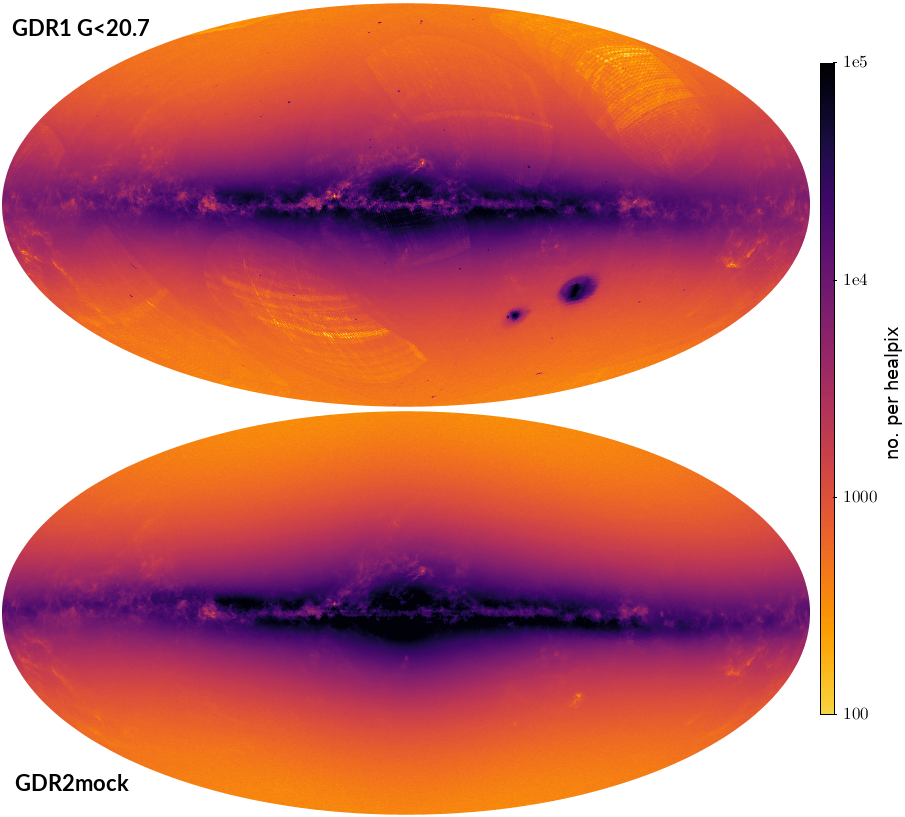}
	\caption{Stellar source density map of GDR1 (top) and our mock catalog (bottom) in Galactic coordinates using Aitoff projection. The Galactic center is in the middle and Galactic longitude $\ell$ increasing towards the left. The color represents the density of star counts down to $G=20.7$\,mag in each healpix (NSIDE = 128, 1\,healpix $\approx0.21\mathrm{deg}^2$) and saturates at both ends to enhance Galactic structures.}	
	\label{fig:density_map}
\end{figure}

\subsection{Dust-attenuated photometry}
\label{sec:extinction}

A crucial step in transforming a \texttt{Galaxia} simulation into a catalog resembling actual observations is the application of a dust distribution,
which will change the apparent colors and luminosities of the stars. 

Since the Gaia photometric bands span a broad wavelength range ($\sim 300$\,nm), the simple conversion of extinction coefficients from e.g. \citet[tab. 6]{2011ApJ...737..103S} to reddening and extinction into the Gaia bands, e.g. A$_G$, is only a poor approximation and may lead to significant inconsistency across the broad range of stellar spectra. Instead we must account for non-linearities in particular with respect to the stars' colors.
Fortunately, the PARSEC isochrones also provide dust attenuated photometry in various photometric systems, including the Gaia passbands (DR1, nominal passbands). 

To include a realistic dust distribution on the \texttt{Galaxia} model, we used the combined 3D extinction map from \citet{2016ApJ...818..130B}, through its python package \texttt{mwdust}\footnote{\url{https://github.com/jobovy/mwdust}}, which is capable of returning line-of-sight extinctions when provided with sky coordinates and distances. This 3D dust map combines the results of \citet{2006A&A...453..635M}, \citet{2015ApJ...810...25G}, and \citet{2003A&A...409..205D} and it provides E(B-V)$_{\mathrm{SFD}}$ values on the scale defined in \citet{1998ApJ...500..525S}\footnote{For a few 3D positions the map returns negative extinctions, but we truncated these to zero.}. As discussed in \cite{2011ApJ...737..103S}, the E(B-V)$_{\mathrm{SFD}}$ scale overestimates the extinction by 14\,\% with respect to their own findings. Hence we corrected for this overestimation and adopted the prescription associated with the PARSEC isochrones of \cite{1989ApJ...345..245C,1994ApJ...422..158O} with R$_0=3.1$ to derive the monochromatic extinction (in mag) at wavelength $\lambda=547.7$\,nm as 
\begin{equation}
\mathrm{A}_0 = 3.1 \times \mathrm{E}(\mathrm{B}-\mathrm{V})_{\mathrm{SFD}} \times 0.86
\end{equation}

\begin{figure}

	\includegraphics[width=\linewidth]{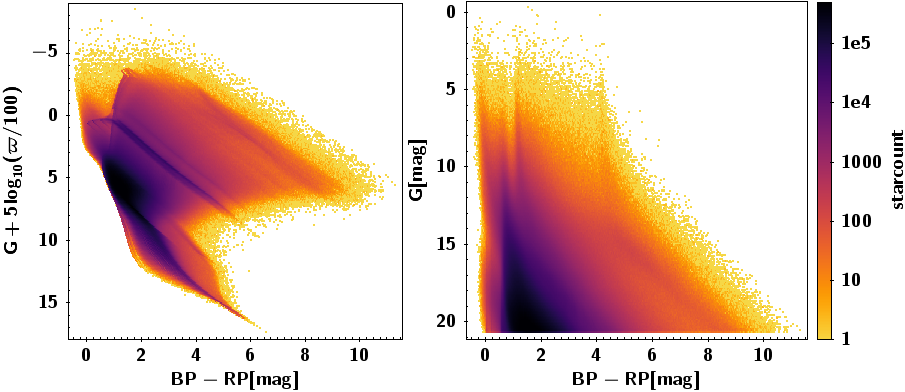}
	\caption{Color-absolute magnitude (left) and color-apparent magnitude (right) diagrams both including extinction in the Gaia photometry system of the 1.6 billion stars in our mock catalog. For every star down to G=20.7\,mag in the \texttt{Galaxia} model we calculated it associated dust attenuated photometry (see Sect.\,\ref{sec:extinction}).  The color for each panel, which represents the stellar density, scales logarithmically. Units are Vega magnitudes with parallax $\varpi$ in mas.}

	\label{fig:CMD}
\end{figure}

Matching each star from \texttt{Galaxia} to an isochrone and a proper amount of extinction is a challenging task for 6 billion stars. 
Instead, we approximated each star to its closest match from a precomputed collection of dust attenuated stellar isochrones.
The grid spans A$_0$ values ranging from 0 to 15\,mag with in steps of 0.025\,mag (for stars with even higher extinction we linearly extrapolated the extinction values) and [Fe/H] values from -2 to 0.5\,dex in steps of 0.25\,dex. We further bin in $\log(\mathrm{T}_\mathrm{eff})$ in 0.02 dex steps and $\log(\mathrm{lum})$ in 0.2 dex steps on a star-by-star basis. Each star in our catalog is associated with an \texttt{index\_parsec} number that records this matching step and maps each star onto the grid of isochrones and thus allows us to query photometric measurements in other bands from the supplementary parsec photometry and extinction table. Figure\,\ref{fig:CMD} shows the resulting color magnitude and absolute magnitude diagrams of the resulting final dataset (applying Gaia selection after accounting for the dust attenuation). 

The following \texttt{ADQL} query provides the data to plot the left panel of Fig.\,\ref{fig:CMD}:
\begin{lstlisting}
SELECT count(*) AS N, 
ROUND(phot_bp_mean_mag - phot_rp_mean_mag, 2) AS color,
ROUND(phot_g_mean_mag + 5 * log10(parallax / 100), 1) AS mag
FROM gdr2mock.main
GROUP BY color,mag
\end{lstlisting}

As the latest PARSEC models (\texttt{v1.2S + COLIBRI}) did not provide dust attenuated photometry
when this catalog was drawn up, we had to match the previous version, \texttt{PARSEC1.2S}\citep{2014MNRAS.444.2525C,2014MNRAS.445.4287T,2015MNRAS.452.1068C} to \texttt{Galaxia}, based on \texttt{PARSEC v1.2S+} isochrones. This inconsistency affects only a limited range of evolution phases that were deeply revised between the two sets of isochrones 
(e.g., O stars, TP-AGB).

\subsection{Additional non-Gaia photometry}

Our catalog provides apparent magnitudes in the nominal DR1 passbands\footnote{The catalog will potentially have an update that uses the GDR2 passbands which will be called gdr2mock\_v2} G, BP, and RP. 
In addition, we provide an additional table, which can be used to obtain photometry for UBVRIJHK \citep{1988PASP..100.1134B,1990PASP..102.1181B,2006AJ....131.1184M}, SDSS \citep{1996AJ....111.1748F}, 2MASS \citep{2003AJ....126.1090C}, and WISE\footnote{\url{http://www.astro.ucla.edu/~wright/WISE/passbaads.html}} \citep{2010AJ....140.1868W} to a precision of $\approx0.1\,\textrm{mag}$. This uncertainty mainly arises from the finite resolution of the isochrone grid we used, which corresponds to  $0.2\,\mathrm{dex}$ spacing in  log-luminosity. With actual GDR2 data, those would be obtained with catalog cross-matching, which of course is not possible with a mock catalog and its random realization of the actual star positions.

The following query illustrates how to obtain complementary photometry (e.g. 2MASS) to the main GDR2mock catalog: 
\begin{lstlisting}
SELECT COUNT(*) AS N, mag_2mass_j AS mag, mag_2mass_j - mag_2mass_ks AS color 
FROM gdr2mock.main AS main
JOIN gdr2mock.photometry AS phot
USING (index_parsec)
WHERE main.random_index <= 1615382
GROUP BY color, mag
\end{lstlisting}
Note that this query also subsamples the catalog to 0.1\,\% using the \texttt{random\_index} and that the queried photometry is in absolute magnitudes. 

\subsection{Uncertainty model}\label{sec:errormodel}

All values provided in the mock catalog are noise-free. 
As a result, there are no negative parallaxes and the parallaxes can be directly inverted to give exact model distances.
To obtain noisy mock observations, one should sample any quantity, say the parallax measurement, from a Gaussian with the \textit{true} \texttt{parallax} as mean and the \texttt{parallax\_error} as the standard deviation. To enable this we provide in the catalog astrometric and photometric uncertainty estimates based on the nominal uncertainty model\footnote{end-of-mission astrometric- and single-transit photometric-uncertainty relations from \url{https://www.cosmos.esa.int/web/gaia/science-performance} requiring V-I color which we calculated internally} \citep{2005ESASP.576...35D} scaled to the duration of the data segment in GDR2 (which is about 668 days or 37\,\% of the 5\,year nominal mission duration). This nominal model depends also on the ecliptic latitude, $\beta$ (which enters via an averaged version of the scanning law). We assume an uncertainty scaling relation of $\frac{1}{\sqrt{n}}$ with the number of observations, $n$, for parallaxes, positions, proper motions and magnitudes, neglecting the noise floors and slightly different scaling for the proper motions based on official communication. 

More specifically, we use an approximation of the Gaia scanning law (scaled to the 22 month data segment) that gives us the number of observations, $n$, as a function of ecliptic latitude in 20 bins\footnote{\url{https://www.cosmos.esa.int/web/gaia/table-2-with-ascii}}. To calculate the parallax uncertainty we use the nominal end-of-mission (eom) parallax uncertainty, $\sigma_{\varpi,\mathrm{eom}}(G,V-I)$, multiply it by the ecliptic latitude dependent uncertainty factor $\mathrm{x}_\varpi(|\sin(\beta)|)$ (\url{https://www.cosmos.esa.int/web/gaia/table-6} which includes the nominal number of observations) and rescale with the shortened 37\,\% baseline:
\begin{equation}
\sigma_{\varpi} = \sigma_{\varpi,\mathrm{eom}}\times  \frac{\mathrm{x}_\varpi}{\sqrt{0.37}}.
\end{equation}
We do the same with the positions and proper motions, which are also related to $\sigma_{\varpi,\mathrm{eom}}$ but have their own ecliptic latitude dependent uncertainty factors provided by the above mentioned online table-6.

For the nominal single-transit (st) photometric uncertainty $\sigma_\mathrm{G,st}(G)$ and $\sigma_\mathrm{BP,RP,st}(G,V-I)$ we simply scale with 1 over the square root of number of observations,
\begin{equation}
\sigma_\mathrm{X} = \frac{\sigma_\mathrm{X,st}}{\sqrt{n}},
\end{equation} 
where X denotes the respective photometric band, i.e. BP, RP, or G.

We do not provide uncertainty estimates for the radial velocity, but the interested reader is referred to \citet{DR2-DPACP-33}.

\subsection{Astrophysical parameters}

A complete simulation of the Milky Way, such as \texttt{Galaxia}, offers not only exact phase-space information of the stars and prediction of their photometric properties, but also of their underlying physical parameters:
ages, masses, metallicities, gravities, luminosities, and effective temperatures, etc. 
These underlying stellar parameters should prove useful in tuning cuts in observables (e.g. color, magnitude and parallax) to optimize for a specific target stellar population (e.g. OB stars, stars with high extinction, old metal-rich stars etc..), and we include them in this mock catalog.
Note that GDR2 will provide observational quantities for some of these stellar parameters, which were derived for sources with $G\le17$\,mag from the Gaia photometry and parallax measurements 
\citep{DR2-DPACP-43}, namely: 
effective temperature for some 161 million sources,  line-of-sight extinction and the reddening, for 88 million sources, and luminosity and radius for 77 million sources.

\section{Catalog Content, Access \& Limitations}

\subsection{Data model and Catalog Content}

Our catalog contains a total number of stars of $1\,606\,747\,036$,
when matching the approximate flux limits of Gaia.
The actual data model of our catalog can be inspected here: \url{http://dc.g-vo.org/tableinfo/gdr2mock.main},
mimicking by design the GDR2 data model\footnote{\url{https://www.cosmos.esa.int/web/gaia/dr2}}: fields and associated names as well as their units. Note, however, that not all columns that appear in DR2 are filled in our catalog and that we provide a few additional ones. Specifically:
\begin{itemize}
	\item \texttt{Nobs} is added, reflecting the nominal ecliptic latitude dependent number of visits for GDR2.
	\item \texttt{age}, \texttt{mass}, \texttt{feh}, \texttt{logg} and \texttt{a0} are added,
    while luminosity, effective temperature, A$_G$, E(BP-RP) and radius are filled into their respective Apsis \citep{2013A&A...559A..74B} fields: \texttt{teff\_val}, \texttt{a\_g\_val}, \texttt{e\_bp\_min\_rp\_val},\texttt{lum\_val} and \texttt{radius\_val}. Beware that in DR2 these are only provided for a subset of stars with $G \leq 17$\,mag (cf. \citealt{DR2-DPACP-43}), whereas in our mock catalog we provide entries for all sources.
	\item \texttt{index\_parsec} is an index for joining the main mock catalog
    to other photometric bands/extinctions in the \texttt{gdr2mock.photometry} table.
\end{itemize}
Similarly to GDR2 we also provide:
\begin{itemize}
\item \texttt{random\_index} is an integer ranging from 0 to $1\,606\,747\,035$, the total number of stars in the mock catalog minus one. This index is useful to create random subsamples representative of the entire catalog. 
    \item \texttt{source\_id} follows the Gaia referencing scheme. It is primarily the healpix\footnote{\url{http://healpix.sf.net}} number using NSIDE = 4096 with the nested scheme in equatorial coordinates multiplied by $2^{35}$. The remaining digits of \texttt{source\_id} are reserved for a running number that serves as a unique identifier per healpix cell. Unlike Gaia no bits are reserved for Data Processing Center identification. Still the \texttt{source\_id} can be easily turned into healpix number for any arbitrary healpix level smaller than 12 (level 12 corresponding to Nside = 4096) via division:
    \begin{equation}
    \mathrm{Healpix}(\mathrm{level}=n) = \frac{\mathtt{source\_id}}{2^{35}\times4^{(12-n)}}
    \end{equation}
\end{itemize}

\subsection{Catalog Access}

The table is available through GAVO's TAP service\footnote{Access URL
\url{http://dc.g-vo.org/tap}, which is also what the run time estimates
refer to} and is registered in the VO registry as\\
\textit{ivo://org.gavo.dc/gdr2mock/q/main}. The full catalog will be hosted by GAVO for at least six month and potentially until GDR3. In the long term there will be a subsample hosted by GAVO which will be cut using the first 10\,\% stars according to the \texttt{random\_index}. However, a bulk download of the complete catalog (without time limitations) is available as FITS binary tables from the reference URL\footnote{\url{http://dc.g-vo.org/tableinfo/gdr2mock.main}}. 

The \texttt{GDR2mock} main table is instantiated using a view (resembling the GDR2 data model) of the actual FITS files. This is why the indexed columns are not marked as such in the \texttt{gdr2mock.main} table but instead in the \texttt{gdr2mock.generated\_data} table. Indexed and therefore fast to query columns are: \texttt{ra}, \texttt{dec}, \texttt{l}, \texttt{b}, \texttt{pmra}, \texttt{pmdec}, \texttt{phot\_g\_mean\_mag}, \texttt{phot\_bp\_mean\_mag}, \texttt{phot\_rp\_mean\_mag}, \texttt{source\_id} and \texttt{random\_index}.

\medskip
It is also planned to host the complete catalog on the Gaia archive (\url{https://gea.esac.esa.int/archive/}).

\subsection{Limitations}
This mock catalog has obvious scientific limitations that stem both from the underlying 
Milky Way model and from our generation of mock observables.

\texttt{Galaxia} is simulating  neither  stellar binaries nor stellar remnants, which will appear in the Gaia data. The phase-space distributions of the stars are assumed smooth and therefore does not generate phase-space or configuration-space clustering. 
The model does not account for extragalactic systems, including LMC, SMC, M31 and M33 which are prominently visible in the GDR1 panel of Figure\,\ref{fig:density_map} and not in our mock catalog. 

To produce extinction estimates, we approximated each star from \texttt{Galaxia} by its nearest model in astrophysical space of a grid of isochrones (see Sect.\,\ref{sec:extinction}).
In addition, observational artifacts were not simulated in our catalog, which can affect the photometry and magnitude limits of stars close to bright sources in the real GDR2 catalog. In particular, we did not attempt to simulate the scanning law and varying magnitude completeness due to crowding issues. 

Finally, this model aims to reproduce the statistical properties of the Milky Way, not its actual properties at the star-by-star level. Hence, cross-matching of our catalog with any other catalog would be moot. 

\section{More example queries}

This catalog offers means to prepare and test ADQL queries for the prospective of GDR2 science cases in a similar run-time environment to the real GDR2 data\footnote{ADQL syntax check on a GDR2 VO service can be run here: \url{http://gaia.ari.uni-heidelberg.de/adql-validator.html}}. Because of the sheer number of sources, a sequential scan (i.e., processing all rows, bypassing indices) will take about an hour wall clock time. This is true for the query on the GAVO service that yielded the data displayed in Figure\,\ref{fig:density_map}: 		
\begin{lstlisting}
SELECT count(*) AS N, ivo_healpix_index(7, ra, dec) AS healpix
FROM gdr2mock.main
GROUP BY healpix
\end{lstlisting}
The real query time may depend on the service used and the server load at the time of query. For more information on the underlying technique, see \citet{2016arXiv161109190T}.

\begin{figure}

	\includegraphics[width=\linewidth]{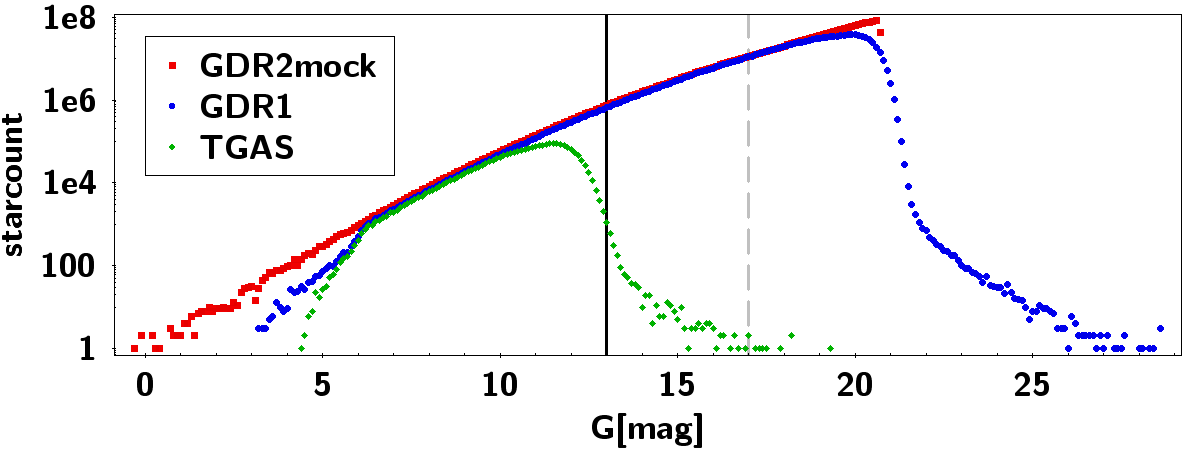}
	\caption{The luminosity function of the GDR2mock catalog in red, compared to Gaia DR1 in blue and its TGAS subset in green. Also indicated are the approximate apparent magnitude limits of the GDR2 radial velocity measurement (solid black) and stellar parameter estimates (dashed gray). The bin size is 0.1\,mag in G.}

	\label{fig:lf}
\end{figure}

We therefore recommend to restrict one's queries to a reasonable spatial subset during the development phase. The \texttt{ADQL} extension Common Table Expressions 
facilitates this. For instance the luminosity function towards the galactic center restricted to a half-degree cone
\begin{lstlisting}
SELECT COUNT(*) AS ct, ROUND(phot_g_mean_mag,1) as bin
FROM gdr2mock.main
WHERE DISTANCE(POINT('GALACTIC', l, b), POINT('GALACTIC', 0., 0.)) < 0.5
GROUP BY bin
ORDER BY bin
\end{lstlisting}
takes just a few seconds and can still run through ``synchronous" query mode, compared to querying the complete luminosity function, which  takes about an hour:
\begin{lstlisting}
SELECT COUNT(*) AS ct, ROUND(phot_g_mean_mag,1) as bin
FROM gdr2mock.main
GROUP BY bin
\end{lstlisting}
For illustration, we compare this last query against the luminosity function of Gaia DR1 and TGAS in Figure\,\ref{fig:lf}.

\begin{figure}

	\includegraphics[width=\linewidth]{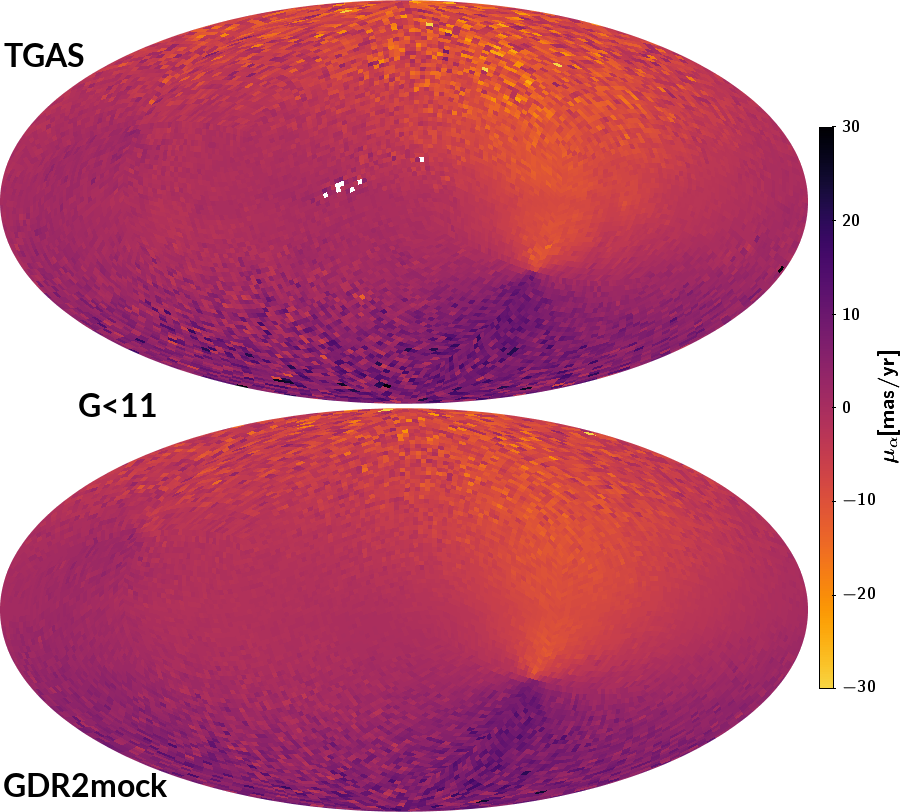}
	\caption{The mean proper motion in right ascension, $\mu_\alpha$, across the sky (Nside = 32, 1 healpix = 3.4 deg$^2$) in Galactic coordinates for TGAS at the top and our mock catalog in the bottom for $G<11$ and $\frac{1}{\varpi}>0.5$\,kpc. The color-coding indicates the mean $\mu_\alpha$ per healpix in mas/yr and saturates at the displayed limits. White pixels in TGAS have no data.}
	\label{fig:proper_motion}
\end{figure}

For bright magnitudes, $G<11$, we can compare the properties of \texttt{GDR2mock} directly with TGAS. For example, we can compare the proper motion in right ascension of TGAS to our catalog. The query for that data is exactly the same for both catalogs except that \texttt{gdr2mock.main} needs to be exchanged for \texttt{tgas.main}:
\begin{lstlisting}
SELECT AVG(pmra) AS mean_pmra, IVO_HEALPIX_INDEX(5, ra, dec) AS healpix
FROM gdr2mock.main
WHERE phot_g_mean_mag < 11 AND 1/parallax > 0.5
GROUP BY healpix
\end{lstlisting}
Figure \ref{fig:proper_motion} shows that overall the two catalogs have similar distributions of motion.

\begin{figure}
	\includegraphics[width=\linewidth]{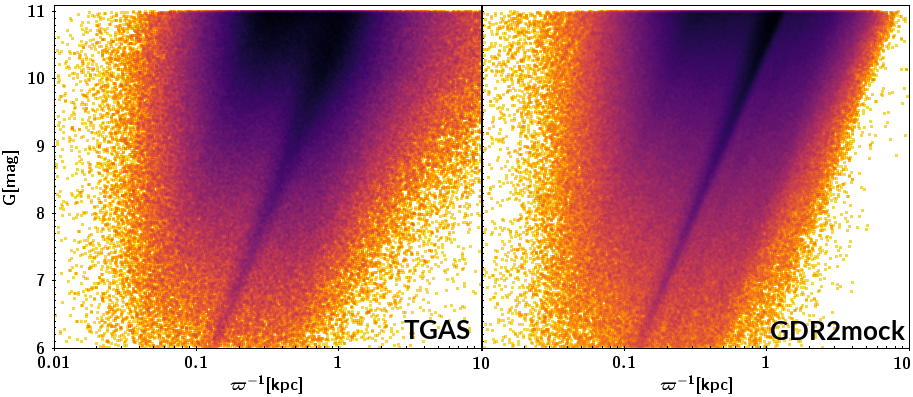}
	\caption{Distance in kpc vs. apparent G magnitude for TGAS (left) and \texttt{GDR2mock} (right). The color-coding shows the log density.}

	\label{fig:distance_g}
\end{figure}

We can also compare the parallaxes between TGAS and \texttt{GDR2mock}. The following query: 
\begin{lstlisting}
SELECT parallax, phot_g_mean_mag
FROM gdr2mock.main
WHERE phot_g_mean_mag < 11
\end{lstlisting}
yields Figure\,\ref{fig:distance_g}, the distribution of stars in (apparent) magnitude -- distance, where the prominent diagonal stripe is composed of red clump stars. 

\begin{figure}
	\includegraphics[width=\linewidth]{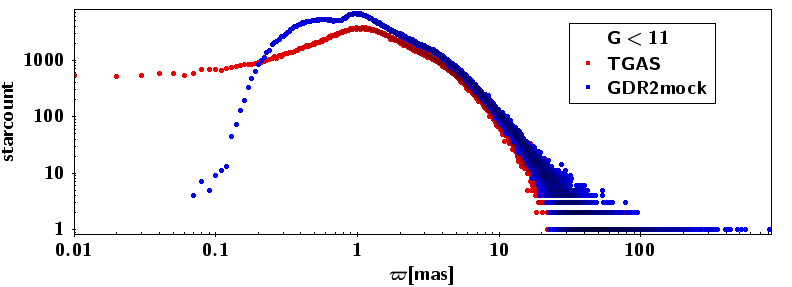}
	\caption{Parallax histogram in 0.01\,mas bins for TGAS and our mock catalog with $G<11$. The tail of negative parallaxes in TGAS is missing in this graphic representation.}
	\label{fig:distance_histogram}
\end{figure}

Similarly, we compare their parallax histograms
\begin{lstlisting}
SELECT COUNT(*) AS ct, ROUND(parallax,2) AS bin
FROM gdr2mock.main
WHERE phot_g_mean_mag<11
GROUP BY bin
\end{lstlisting}
in Figure\,\ref{fig:distance_histogram}, which illustrates the difference between true (\texttt{GDR2mock}) and measured (TGAS) parallaxes (i.e. inclusion of measurement uncertainties). Beware that parallax measurements from GDR2 will be more accurate than from TGAS, even though the nominal uncertainty model is very optimistic for those bright stars. When for example sampling \textit{observed} parallaxes for this $G<11$ subsample of \texttt{GDR2mock} using \texttt{parallax} and \texttt{parallax\_error} the chances of measuring a non-positive parallax at all is below 1\,\%.

\begin{figure}
	\includegraphics[width=\linewidth]{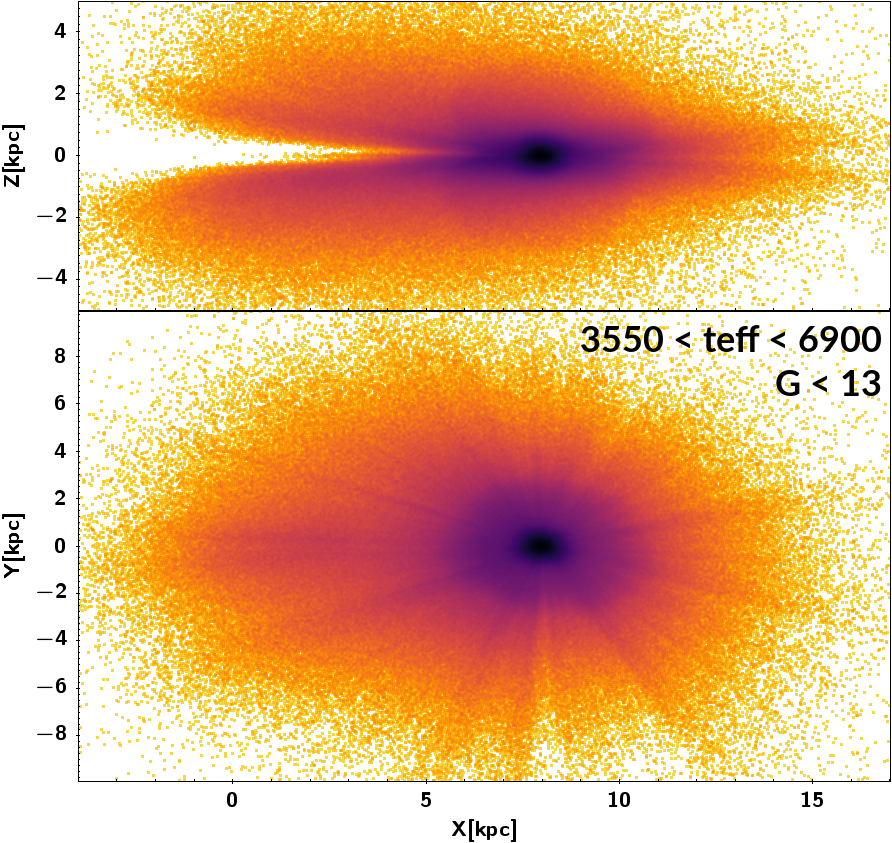}
	\caption{Spatial distribution of stars in the GDR2mock catalog with $G<13$ and $3550<\texttt{teff\_val}<6900$, illustrating the expected volume for which GDR2 will provide full 6D phase-space information. The color encodes logarithmic density. The Fingers of God effect is due to dust along the line of sight, the observer being centered at solar position i.e. $(X,Y,Z) = (8.0,0.0,0.015)\,\mathrm{kpc}$}
	\label{fig:xyz}
\end{figure}

The distribution of stars in the Galaxy which will have radial velocities in GDR2 is displayed in Figure\,\ref{fig:xyz}, which resulted from the query
\begin{lstlisting}
SELECT 8 - COS(RADIANS(b)) * (1/parallax) * COS(RADIANS(l)) AS x, COS(RADIANS(b)) * (1/parallax) * SIN(RADIANS(l)) AS y, 0.015 + (1/parallax) * SIN(RADIANS(b)) AS z
FROM gdr2mock.main
WHERE phot_g_mean_mag < 13 AND teff_val > 3550 AND teff_val < 6900
\end{lstlisting}

\section{Summary}
We presented a simulation of the Gaia DR2 stellar content which can be accessed via \url{http://dc.g-vo.org/tableinfo/gdr2mock.main}. Using \texttt{Galaxia} and realistic 3D extinction maps we have produced a catalog, GDR2mock, that closely resembles the Gaia observations (cf. Figure\,\ref{fig:density_map}). Together with the scaled nominal uncertainty estimates, our mock catalog will give the scientific community a convenient tool to hone queries and know what to 
expect from GDR2; beyond the GDR2 release, this mock catalog provides a valuable comparison for science analysis. It should serve as a test-bed for first day GDR2 scientific projects (in runtime and \texttt{ADQL} syntax), as well as a comparison to real queries in order to establish field contamination or confirm unexpected features.

\section{Acknowledgements}
We thank the anonymous referee for their prompt report. The authors thank Leo Girardi, Jo Bovy, Sanjib Sharma and Alcione Mora for their useful help.

This work made use of \texttt{topcat} \citep{2005ASPC..347...29T}, \texttt{HEALPix} \citep{2005ApJ...622..759G}, \texttt{astropy}, and \texttt{ezpadova}\footnote{\url{https://github.com/mfouesneau/ezpadova}} suites and packages. 

We thank the German Astrophysical Virtual Observatory\footnote{\url{http://www.g-vo.org/}} for the publishing platform and for fruitful discussions on the technical
aspects of this endeavor. 

This work was funded in part by the DLR (German space agency) via grant 50\,QG\,1403. 
J.R and H.W.R. acknowledge funding from the
European Research Council under the European Union’s Seventh Framework
Programme (FP 7) ERC Advanced Grant Agreement No. [321035]

This project was developed in part at the 2017 Heidelberg Gaia Sprint, hosted by the Max-Planck-Institut f{\"u}r Astronomie, Heidelberg.

This work has made use of data from the European Space Agency (ESA) mission Gaia, processed by the Gaia Data Processing and Analysis Consortium (DPAC). Funding for the DPAC has been provided by national institutions, in particular the institutions participating in the Gaia Multilateral Agreement.

\bibliographystyle{aasjournal} 
\bibliography{adslib,otherlib} 
\end{document}